\documentclass[journal]{IEEEtran}

\ifCLASSINFOpdf
\else
\fi
\usepackage{cite}
\usepackage[cmex10]{amsmath}

\newcommand{\vect}{\operatorname{vec}}
\usepackage{amssymb}   % \triangleq vs. gibi seyler bunda
\usepackage{amsxtra}
\usepackage{amscd}
\usepackage{amsthm}
\usepackage{amsxtra}
\usepackage{amscd}
\usepackage{amsthm}
\usepackage{tikz}
\usetikzlibrary{matrix}
\usepackage{mathtools}
\usepackage{graphicx}   % sekilller icin
\usepackage{wrapfig}
\usepackage{layouts}
\usepackage{array}  % tablo icin satir acma
\usepackage{multirow}  % tablolar icin coklu satir kullanmada
\usepackage{arydshln}
\usepackage{mathtools}
\usepackage{xparse}% http://ctan.org/pkg/xparse
\usepackage[caption=false]{subfig}
\DeclareMathOperator{\diag}{diag}
\NewDocumentCommand{\overarrow}{O{=} O{\uparrow} m}{%
	\overset{\makebox[0pt]{\begin{tabular}{@{}c@{}}#3\\[0pt]\ensuremath{#2}\end{tabular}}}{#1}
}
\NewDocumentCommand{\underarrow}{O{=} O{\downarrow} m}{%
	\underset{\makebox[0pt]{\begin{tabular}{@{}c@{}}\ensuremath{#2}\\[0pt]#3\end{tabular}}}{#1}
}

\usepackage{verbatim}
\usepackage{stfloats}
\usepackage{blindtext}
\usepackage{abraces}
\usepackage{algpseudocode}
\usepackage{algorithm}
\usepackage{color}

\usepackage{xparse}
\usepackage{cite}
\renewcommand{\thefootnote}{\arabic{footnote}}
\usepackage[keeplastbox]{flushend}

\newcommand\blfootnote[1]{%
	\begingroup
	\renewcommand\thefootnote{}\footnote{#1}%
	\addtocounter{footnote}{-1}%
	\endgroup
}

\usetikzlibrary{decorations.pathreplacing,
	matrix,
	positioning,
}
\usepackage[bottom]{footmisc}
\usepackage{tikz}

\usetikzlibrary{decorations.pathreplacing}
\newcommand{\vast}{\bBigg@{4}}
\newcommand{\Vast}{\bBigg@{5}}
\usepackage{float}
\usepackage{xcolor,cite,etoolbox}

\begin{document}
	\bstctlcite{IEEEexample:BSTcontrol}
	\setlength{\abovedisplayskip}{3.5pt}
	\setlength{\belowdisplayskip}{3.5pt}
\title{Low Complexity Adaptation for Reconfigurable Intelligent Surface-Based MIMO Systems}

\author{Zehra~Yigit,~\IEEEmembership{Student~ Member,~IEEE,}
        Ertugrul~Basar,~\IEEEmembership{Senior~Member,~IEEE,}
        and~Ibrahim~Altunbas,~\IEEEmembership{Senior~Member,~IEEE}% <-this % stops a space
\thanks{Z. Yigit and Ibrahim Altunbas are with the Department of Electronics and Communication Engineering, Istanbul Technical University,  Maslak 34469, Istanbul, Turkey. E-mail: {yigitz@itu.edu.tr, ibraltunbas@itu.edu.tr}.}% <-this % stops a space
\thanks{E. Basar is with the Communications Research and Innovation Laboratory (\text{CoreLab}),  Department of Electrical and Electronics Engineering, Ko\c{c} University, Sariyer 34450, Istanbul, Turkey. E-mail: ebasar@ku.edu.tr}.}

\markboth{IEEE COMMUNICATIONS LETTERS, VOL. XX, NO. X, 2020}
{Shell \MakeLowercase{\textit{et al.}}: Bare Demo of IEEEtran.cls for IEEE Journals}
\maketitle

\begin{abstract}
Reconfigurable intelligent surface (RIS)-based transmission technology offers a promising solution to enhance wireless communication performance cost-effectively through properly adjusting the parameters of a large number of passive reflecting elements.  
This letter proposes a cosine similarity theorem-based low-complexity algorithm for adapting the phase shifts of  an RIS  that assists a multiple-input multiple-output (MIMO) transmission system.  A semi-analytical probabilistic approach is developed to derive the theoretical average {bit} error probability (ABEP) {of the system}. Furthermore,  the validity of the theoretical analysis is supported through extensive computer simulations.
\end{abstract}

\blfootnote{This work was supported by the Scientific and Technological Research Council of Turkey (T{\"U}B{\.I}TAK) under grant no 117E869.}

% \fi
\IEEEpeerreviewmaketitle

\vspace*{-0.45cm}
\section{Introduction}
\IEEEPARstart{T}{he} fifth generation (5G) wireless communication technology promises  an explosive growth on data rate, massive connectivity and latency performance. To achieve these goals, various transmission technologies have been developed in  recent years. Massive multiple-input multiple-output (MIMO) and milimeter wave (mmWave) communication systems are considered as some of the prominent candidates among these technologies. On the other hand, to meet this challenge beyond 5G requirements, utilization of an increasing number of multi-antenna systems has raised  strong concerns  about the energy efficiency and hardware cost of large-scale MIMO systems.   

Recently, reconfigurable intelligent surface (RIS)-assisted communication  technology has been considered as a promising solution to overcome  the energy efficiency related issues of future wireless networks \cite{zhang2019capacity,nadeem2019large, basar2019transmission}. An RIS is a planar metasurface that consists of a number of low-cost passive reflecting elements, each of which smartly induces an independent phase shift to modify the propagation environment in more favorable way for the communication performance \cite{basar2019wireless}. 

The unprecedented potential of  RISs on the signal quality  of a communication system has led researchers to largely consider the  RIS technology in  various frontiers. In  one of the early studies \cite{basar2019transmission}, the error performance of an  RIS aided single-input single-output (SISO) system is investigated through a mathematical framework. Also,  in \cite{basar2020reconfigurable, gopi2020intelligent}, to improve the spectral efficiency, the RIS  aided index modulation (IM)  systems have been proposed. 
Later,  RISs are considered for  multiuser systems to enhance  the energy efficiency  \cite{huang2019reconfigurable} and to maximize the signal-interference-noise ratio (SINR)  \cite{nadeem2019large, wu2019intelligent, hou2019mimo}. Even more recently, {to facilitate practical implementation of the RIS aided systems,} %RIS-assisted multi-antenna systems,
 new  path loss models \cite{ellingson2019path,tang2019wireless}, { open-source channel models \cite{ basar2020indoor,basar2020simris} and  a practical phase shift model \cite{abeywickrama2020intelligent}  have been developed}.

%Even more recently, for the RIS assited multi-antenna systems, new channel estimation protocols \cite{nadeem2019intelligent, he2020channel} and path loss models  \cite{bjornson2020power,ellingson2019path,tang2019wireless} have been developed. Also,  in \cite{basar2020reconfigurable}, to increase spectral efficiency, a RIS aided index modulation (IM)  system has been proposed. 

%Notably, the most of the existing RIS-aided designs are based on computationally intensive and complex algorithms \cite{nadeem2019intelligent, huang2019reconfigurable,hou2019mimo, nadeem2019large} and RIS-aided designs have been carried out for  SISO and MISO systems \cite{basar2019transmission, basar2020reconfigurable,nadeem2019intelligent, huang2019reconfigurable}. 
Notably,  most of the existing  RIS-aided  designs  developed for  multiple-input single-output (MISO) \cite{wu2019intelligent, huang2019reconfigurable,nadeem2019large, wu2019weighted} and MIMO systems  \cite{zhang2019capacity, hou2019mimo}  exploit   computationally intensive and complex algorithms.   Specifically, no single study exists  performing  a statistical analysis on the theoretical bit error rate (BER) performance of RIS-aided multi-antenna systems. 

In this letter, an efficient  low-complexity algorithm,  which is based on cosine similarity theorem \cite{moon2000mathematical} and  adapts 
the phase shift of each reflecting elements, is proposed for 
%improving  the overall path gains of   
RIS-aided MIMO  (RIS-MIMO) and RIS-aided spatial modulation (RIS-SM) schemes in which the transmission principles of classical MIMO and classical SM \cite{mesleh2008spatial,basar2020reconfigurable} are considered, respectively.  Moreover, a semi-analytical probabilistic model is developed to derive the average bit error probability  (ABEP) of the proposed RIS-MIMO and RIS-SM schemes. %Furthermore, through computer simulations the validity of our semi-analytic model is approved.
 { Furthermore, through the comprehensive computer simulations, the BER performance of the RIS-aided MIMO systems with the proposed algorithm is investigated for the various conditions including perfect/imperfect channel estimation, continous/discrete phase reflection and path loss effect. }
%The error performance of the RIS-MIMO and RIS-SM schemes are investigated through semi-analytic  and  algorithm

The rest of this letter is structured as follows. In Section II, the system model of the proposed RIS-aided MIMO  transmission systems is presented. The theoretical performance analysis of the  system is carried out in Section III. Computer simulation results are described in Section IV and the paper is concluded in Section V\footnote{\textit{Notation:} Bold  lowercase and uppercase letters are used for vectors and matrices, respectively. $(\cdot)^\mathrm{T}$, $(\cdot)^\mathrm{H}$ and \mbox{$||\cdot||$} stand for transposition, Hermitian transposition and Euclidean/Frobenious norm operators, respectively. $\vect(\cdot)$ denotes vectorization operator and $\det(\cdot)$ symbolizes determinant of a matrix. The Kronecker product    and  Euclidean inner product of two vectors are denoted by  $\otimes$ and $<\cdot,\cdot>$, respectively. $\mathbb{C}^{a\times b} 
	$ represents the set of $a\times b$ dimensional matrices while  $\diag\left\lbrace \cdot\right\rbrace $ symbolizes a square diagonal matrix. $\mathcal{CN}(\mu,\sigma^2)$ denotes the complex Gaussian distribution of a random variable with $\mu$ mean and $\sigma^2$ variance. $\mathcal{O}(\cdot)$ and $P(\cdot)$ stand for the big $\mathcal{O}$ notation and probability of an event, respectively. $\mathbf{I}_n$ denotes $n\times n$  identity matrix while $\mathbf{1}$  represents all-ones column vector.} . 
		\vspace*{-0.3cm}
\section{System Model}
In this section, the concept of the proposed low-complexity algorithm and the system models of the RIS-MIMO and RIS-SM schemes are  introduced.    In the proposed systems, the transmitter and the receiver are assumed to be equipped with  $T_x$ and $R_x$ antennas, respectively, as shown \mbox{in Fig. 1.}  In addition, an RIS with $N$ passive reflecting elements {is used to improve the communication performance by appropriately adjusting phase shift of each reflecting element.}  
\begin{figure}[t]	
	\centering
	\includegraphics[width=0.85\linewidth]{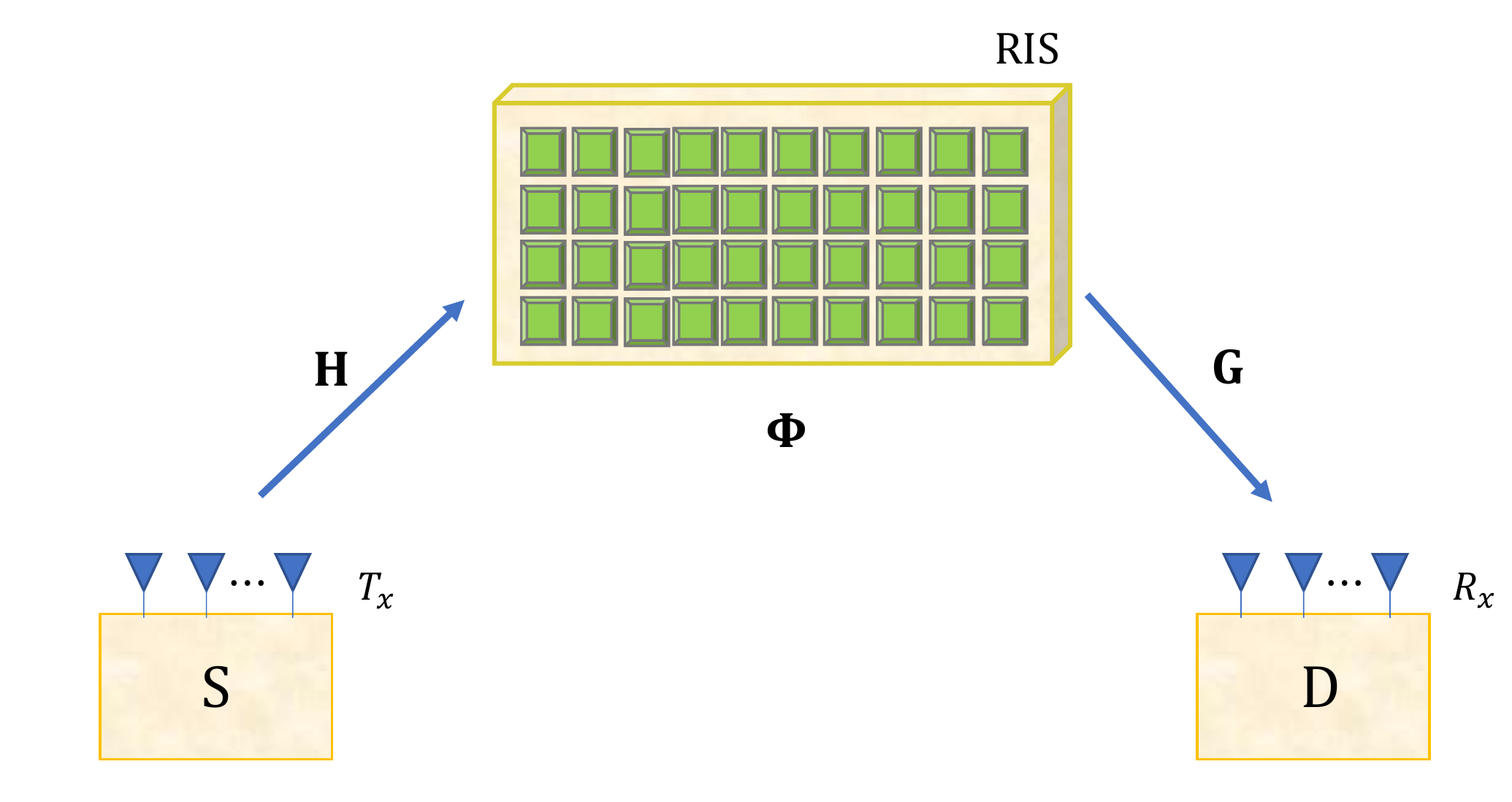}
			\vspace*{-0.1cm}
	\caption{RIS aided MIMO Systems}
	\label{sys}
			\vspace*{-0.1cm}
\end{figure}
Consider $\mathbf{H}\in\mathbb{C}^{N\times T_x}$ and $\mathbf{G}\in\mathbb{C}^{N\times R_x}$ as the matrices of uncorrelated Rayleigh fading channel  from the transmitter to the RIS, and from  the RIS to the receiver, respectively, whose  elements are independent and identically distributed (i.i.d.) and follow   $\mathcal{CN}(0,1)$ distribution. On the other hand,  $\mathbf{\Phi}\in\mathbb{C}^{N\times N}$ stands for the  matrix of  RIS reflection coefficients with {$\mathbf{\Phi}=\diag\left\lbrace \beta_1e^{j\phi_1}, \beta_2e^{j\phi_2}, \ldots, \beta_Ne^{j\phi_N}\right\rbrace $}, where $\phi_i\in[-\pi, \pi]$ is  phase shift and {$\beta_i\in(0,1]$ is the amplitude reflection coefficient} of the $i$th reflecting element, for $i\in\left\lbrace 1,2,\ldots, N\right\rbrace $. {However, for the sake of simplicity, we set $\beta_i=1$.} Therefore, the composite MIMO channel matrix $\mathbf{C}\in\mathbb{C}^{R_x\times T_x}$ from the transmitter to the receiver becomes $\mathbf{C}=\mathbf{G}^\mathrm{H}\mathbf{\Phi H}$.  In the proposed systems,  perfect channel state information (CSI) of all channels are available at all nodes and quasi-static block fading channels are assumed.
\vspace*{-0.05cm}
\subsection{Proposed Algorithm}
%{\color{red}In this paper, it is aimed to maximize the total channel gain of the RIS aided system by arranging the each  phase of the reflecting elements through a cos of the considered RIS. }
%According to Cauchy Schwarz inequality \cite{moon2000mathematical},  since $||\mathbf{\Phi}||$ is unitary, the maximum achievable gain of the composite channel $\mathbf{C}$ is calculated by:
{In this subsection, we  develop  a low-complexity algorithm to maximize  the average received signal-noise ratio (SNR) of the  RIS aided MIMO systems, which results in maximizing the overall channel gain of the system by arranging the phase shift
	of each reflecting element.
	{Then}, using $\mathbf{C}=\mathbf{G}^\mathrm{H}\mathbf{\Phi H}=\sum_{i=1}^{N}\mathbf{g}_i^He^{j{\phi}_i}\mathbf{h}_i$, our  problem is formulated  as 
	\begin{align}
	&\vspace*{-0.3cm}\max_{{\phi}_i} 
	\left\|\mathbf{C}\right\|=\max_{{\phi}_i}||\sum\nolimits_{i=1}^{N}\mathbf{g}_i^He^{j{\phi}_i}\mathbf{h}_i||\nonumber\\%\leq \sum_{i=1}^{N}| |\mathbf{g}_i^\mathrm{H}|| \left\| \mathbf{h}_i\right\|
	&\hspace{0.1cm}\mathrm{s.t.}\hspace{0.5cm}  |e^{j\phi_i}|=1
	\label{cs2}
	\end{align}
	where $\mathbf{g}_i^\mathrm{H}$  and $\mathbf{h}_i$ stand for  the $i$th column and $i$th row of  $\mathbf{G}^\mathrm{H}$  and $\mathbf{H}$, respectively. %The maximization problem in (\ref{cs2})  is non-convex due to the constraint $\phi_i\in[-\pi,\pi]$. 
	Although the maximization problem in (\ref{cs2})  is non-convex due to the constraint $\phi_i\in[-\pi,\pi]$, the achievable channel gain can be upper bounded as
	\begin{equation}
	\left\|\mathbf{C}\right\|=||\sum\nolimits_{i=1}^{N}\mathbf{g}_i^He^{j{\phi}_i}\mathbf{h}_i||\leq \sum\nolimits_{i=1}^{N}| |\mathbf{g}_i^\mathrm{H}|| \left\| \mathbf{h}_i\right\|.
	\label{cs22}
	\end{equation}
	Then, exploiting (\ref{cs22}),  the maximum achievable gain of the component at the  $k$th row and the $l$th column of   $\mathbf{C}$, shown by $c_{k,l}$,  can be given as
	%according to the Cauchy Schwarz inequality \cite{moon2000mathematical},  the maximum achievable gain of $c_{k,l}$ component at the  $k$th row and the $l$th column of  the $\mathbf{C}$ matrix  can be given as
	\begin{equation}
	%\left\|\mathbf{C}\right\|=||\sum_{i=1}^{N}\mathbf{g}_i^He^{j{\phi}_i}\mathbf{h}_i||\leq \sum_{i=1}^{N}| |\mathbf{g}_i^\mathrm{H}|| \left\| \mathbf{h}_i\right\|\\
	\left| c_{k,l}\right|  =| \sum\nolimits_{i=1}^Ng_{k,i} e^{j\phi_i}h_{i,l}|\leq \sum\nolimits_{i=1}^N|g_{k,i}||h_{i,l}|
	\label{cs3}
	\end{equation} where $g_{k,i}$ and $h_{i,l}$ are the $k$th and $l$th components of the channel vectors $\mathbf{g}_i^\mathrm{H}$  and  $\mathbf{h}_i$, respectively, for $k\in\left\lbrace 1,2,\ldots,R_x\right\rbrace $ and $l\in\left\lbrace 1,2,\ldots,T_x\right\rbrace $. It is quite obvious that  there exists  an optimum  $\mathbf{\Phi}$ matrix that satisfies (\ref{cs3}) with an equality to achieve  the maximum channel gain.  {However, due to the non-convex constraints, it is difficult to find  an optimal solution} {for this problem in a computationally efficient and robust manner. Moreover, this challenge increases when a multi-antenna system is considered at both the transmitter and the receiver.}  %Therefore,it is not easy to find  an optimal solution for this  in a computationally efficient and robust manner%{\color{red}, which has led researchers to develop suboptimal solutions}
	Due to these limitations, we develop an efficient suboptimal solution using the  {cosine similarity theorem \cite{moon2000mathematical}, where the  angle between   $\mathbf{u}$ and $\mathbf{v}$ vectors   is calculated, using their inner product and magnitudes
		\cite{moon2000mathematical}, as 
		\begin{equation}
		\cos(\mathbf{u}, \mathbf{v})=\frac{<\mathbf{u},\mathbf{v}>}{||\mathbf{u}||||\mathbf{v}||}.
		\end{equation}}% to achieve the maximum channel gain of the overall system with low %, using the cosine similarity theorem \cite{moon2000mathematical}, 
	
	%the phase shift $\phi_i$ is adjusted to approximate the complex channel vectors  $\mathbf{h}_i$ and $\mathbf{g}_i^\mathrm{H}$  to their component-wise absolute vectors $\tilde{\mathbf{h}}_i=[|h_{i,1}|,|h_{i,2}|, ,\ldots|h_{i,T_x}|]$     and $\tilde{\mathbf{g}}_i^\mathrm{H}=[|g_{1,i}|,|g_{2,i}|, ,\ldots|g_{R_x,i}|]$, respectively. 
	In the proposed algorithm,  to maximize the individual  gain of  each $c_{k,l}$ component   (\ref{cs3})   the phase shift $\phi_i$  is adjusted to approximate the complex channel vectors  $\mathbf{h}_i$ and $\mathbf{g}_i^\mathrm{H}$  to their component-wise absolute vectors $\tilde{\mathbf{h}}_i=[|h_{i,1}|,|h_{i,2}|, ,\ldots|h_{i,T_x}|]$     and $\tilde{\mathbf{g}}_i^\mathrm{H}=[|g_{1,i}|,|g_{2,i}|, ,\ldots|g_{R_x,i}|]^\mathrm{H}$, respectively. 
	Therefore, in  \mbox{Algorithm 1},    real $\phi_i^h$ and  $\phi_i^g $   angles are calculated  in order to measure the  cosine similarity between the vectors $\mathbf{h}_i$ and $\tilde{\mathbf{h}}_i$ and   $\mathbf{g}_i^\mathrm{H}$ and $\tilde{\mathbf{g}}_i^\mathrm{H}$, respectively.  Then, for the $i$th reflecting element, the  overall phase shift $\phi_i$ is determined as $\phi_i=-(\phi_i^h+\phi_i^g)$ and the overall reflection matrix $\mathbf{\Phi}$ is obtained  accordingly.
	\begin{algorithm}[t]
		\caption{ Cosine Similarity-based Low-Complexity Algorithm}
		\hspace*{\algorithmicindent} \textbf{Input}:  ${\mathbf{h}}_i$, $\mathbf{g}_i$, $\tilde{\mathbf{h}}_i$, $\tilde{\mathbf{g}}_i$ \\
		\hspace*{\algorithmicindent} \textbf{Output}: $\mathbf{\Phi}$ 
		\begin{algorithmic}	[1]
			\For{$i = 1:N$}
			\State $\phi_i^h=\arccos\left(  \frac{<\mathbf{h}_i, \tilde{\mathbf{h}}_i>_{\operatorname{Re}}}{| |\mathbf{h}_i|||| \tilde{\mathbf{h}}_i||} \right) $
			\State $\phi_i^g=\arccos\left(  \frac{<\mathbf{g}_i^\mathrm{H}, \tilde{\mathbf{g}}_i^\mathrm{H}>_{\operatorname{Re}}}{|| \mathbf{g}_i^\mathrm{H}|||| \tilde{\mathbf{g}}_i^\mathrm{H}||} \right) $
			\State $\phi_i=-(\phi_i^h+\phi_i^g)$
			\EndFor
			\State $\mathbf{\Phi}=\diag(e^{j\phi_1}, e^{j\phi_2}, \ldots, e^{j\phi_N} )$
		\end{algorithmic}
			\vspace*{-0.1cm}
	\end{algorithm} 
	\setlength{\textfloatsep}{11pt}
	
	Let us present the concept of the proposed algorithm  by an example. Consider an RIS-aided MIMO system, with 
	$T_x=2$, $R_x=3$ ve $N=2$ whose composite channel matrix  $\mathbf{C}=\mathbf{G}^\mathrm{H}\mathbf{\Phi H}$ is constructed as
	\begin{align}
	&\mathbf{C}=\begin{bmatrix}
	g_{11}&g_{12}\\
	g_{21}&g_{22}\\
	g_{31}&g_{32} 
	\end{bmatrix}\begin{bmatrix}
	e^{j\phi_{1}}&0\\
	0&e^{j\phi_{2}}\\
	\end{bmatrix}
	\begin{bmatrix}
	h_{11}&h_{12}\\
	h_{21}&h_{22}
	\end{bmatrix}
	\label{cm1} 
	\end{align} 
	which can be rewritten, in the form of $\mathbf{C}=\sum_{i=1}^{N}\mathbf{g}_i^He^{j{\phi}_i}\mathbf{h}_i $, as% given for $i$th   $\mathbf{g}_i^\mathrm{H}$  and $\mathbf{h}_i$ as
	\begin{align}
	&\mathbf{C}=\begin{bmatrix}
	g_{11}\\
	g_{21}\\
	g_{31}
	\end{bmatrix}e^{j\phi_{1}}\begin{bmatrix}
	h_{11}&h_{12}
	\end{bmatrix}+\begin{bmatrix}
	g_{12}\\
	g_{22}\\
	g_{32}
	\end{bmatrix}e^{j\phi_{2}}\begin{bmatrix}
	h_{21}&h_{22}\end{bmatrix}.
	\label{c2}
	\end{align}
	Since the phase shift  $\phi_i$  only affects  $\mathbf{g}_i^\mathrm{H}$  and $\mathbf{h}_i$  vectors,  instead of  {jointly adjusting} all phases,
	each   $\phi_{i}$  can be individually determined in a  more computationally efficient manner. Therefore, {in the proposed algorithm},  considering (\ref{cs3}),  we singly determine  each $\phi_i$   to improve the overall channel gain. For this aim, $\phi_i$ is properly adjusted to make   the channel vectors $\mathbf{h}_i$ and $\mathbf{g}_i^\mathrm{H}$  approximate to their component-wise absolute  vectors $\tilde{\mathbf{h}}_i=[|h_{i,1}|,|h_{i,2}|]$, and  $\tilde{\mathbf{g}}_i^\mathrm{H}=[|g_{1,i}|,|g_{2,i}|, |g_{3,i}|]^\mathrm{H}$, respectively.   Therefore,  as given in Algorithm 1, %since $\phi_i\in[-\pi,\pi]$, 
	the real  angles $\phi_i^h$ and $\phi_i^g$, which respectively measure the similarities between  the  vectors  $\mathbf{h}_i$ and $\tilde{\mathbf{h}}_i$, and  the vectors $\mathbf{g}_i^\mathrm{H}$  and    $\tilde{\mathbf{g}}_i^\mathrm{H}$, are   calculated using the cosine similarity theorem \cite{moon2000mathematical}.
	%\cite{moon2000mathematical}
	%\begin{equation}
	% \phi_i^h=\arccos\left(  \frac{<\mathbf{h}_i, \tilde{\mathbf{h}}_i>_{\operatorname{Re}}}{| |\mathbf{h}_i|||| \tilde{\mathbf{h}}_i||} \right),\hspace{0.1cm}\phi_i^g=\arccos\left(  \frac{<\mathbf{g}_i, \tilde{\mathbf{g}}_i>_{\operatorname{Re}}}{| |\mathbf{g}_i|||| \tilde{\mathbf{g}}_i||} \right)
	% \end{equation}
	Then, the effective phase shift of $i$th reflecting elements is set to $\phi_i=-(\phi_i^h+\phi_i^g)$. After $N$ repetitions are performed, the overall 
	reflection matrix $\mathbf{\Phi}$ is constructed.
	
	The required complexity to perform this algorithm is $\mathcal{O}(6N(T_x+R_x))$ in terms of real multiplications (RMs). 
	It is worth noting that  when a SISO system  ($T_x=R_x=1$) is considered, this algorithm satisfies (\ref{cs3}) with an equality that  achieves the optimum channel gain  in \cite{basar2019transmission}.
	\vspace{-0.5cm} 
	\subsection{RIS-MIMO Scheme}
	In the proposed RIS-MIMO scheme with  $T_x$ transmit antennas, $R_x$ receive antennas and  $N$ reflecting elements,    a classical MIMO  transmission principle is applied and  the reflection parameters are determined by performing the proposed low-complexity algorithm to improve the  overall channel gain.  %whose reflection parameters are determined through the proposed cosine similarity theorem-based algorithm, 
	Let $\mathbf{x}=\left[ x_1, x_2, \ldots, x_{T_x}\right] ^\mathrm{T}\in\mathbb{C}^{T_x\times 1}$ be  the transmitted signal vector of the RIS-MIMO scheme, %which is simultaneously transmitted through available $T_x$ transmit antennas
	where $x_j$ denotes an $M$-ary phase shift keying/quadrature amplitude modulation (PSK/QAM) symbol transmitted through $j$th transmit antenna, for $j\in\left\lbrace1,2,\ldots, T_x \right\rbrace 
	$. %Therefore, the achieved spectral efficiency of the proposed RIS-MIMO system is
	%\begin{equation}
	%m=T_x\log2(M)\hspace{0.5cm}\text{[bits/s/Hz]}.
	%\label{eq0}
	%\end{equation}
	Then,  the received signal vector $\mathbf{y}\in\mathbb{C}^{R_x\times 1}$ becomes
	\begin{align}
	&\mathbf{y}=\mathbf{G}^\mathrm{H}\mathbf{\Phi Hx}+\mathbf{n}\nonumber\\
	&\hspace{0.25cm}=\mathbf{Cx}+\mathbf{n}
	\label{eq.1}
	\end{align}
	where $\mathbf{n}\in\mathbb{C}^{R_x\times 1}$ is the vector of additive white Gaussian noise samples whose elements are i.i.d. and  follow $\mathcal{CN}(0,N_0)$ distribution. %In the  RIS-MIMO scheme, the effective reflection matrix $\mathbf{\Phi}$  is determined by performing the proposed low-complexity algorithm to improve the overall channel gain. 
	\vspace*{-0.3cm}
	\subsection{RIS-SM Scheme}
	In this subsection, the system model of the RIS-SM scheme,  where  an RIS-aided MIMO  scheme applying the traditional SM transmission principle \cite{mesleh2008spatial}  is presented.  Unlike traditional SM \cite{mesleh2008spatial}, in the proposed RIS-SM scheme, the transmitted signal quality is significantly improved by the RIS, whose each reflection parameter is arranged using the proposed low-complexity algorithm. 
	
	In RIS-SM,  one of the $T_x$ transmit antenna is activated and a modulated symbol $s$ from the $M$-PSK constellation is  transmitted through this active antenna.  Since the remaining transmit antennas are deactivated, the signal vector is determined as $\mathbf{x}=\begin{bmatrix}
	0&\ldots0&s&0&\ldots&0
	\end{bmatrix}^\mathrm{T}$ and transmitted through the overall channel matrix $\mathbf{C}$ as given in (\ref{eq.1}). 
	
	{Note that in the proposed RIS-SM scheme,  the knowledge of   the active antenna index is not provided to the RIS.}
	
	At the receiver of both RIS-MIMO and RIS-SM schemes, to obtain the best BER performance, a maximum likelihood (ML) detector is considered, and the transmit RIS-SM and RIS-MIMO vectors are detected by considering all possible $\mathbf{x}$ realizations as follows
	\begin{equation}
	\hat{\mathbf{x}}=\arg \min_{\mathbf{x}}\left\| \mathbf{y}-\mathbf{Cx}\right\|^2. 
	\label{ml1}
	\end{equation}
	\section{Performance Analysis}
	In this section, the theoretical BER performance of the proposed RIS-MIMO and RIS-SM schemes is evaluated by performing a semi-analytical probabilistic  approach.  %Since the proposed  algorithm relate the phase of each reflecting element with its corresponding arriving and leaving channel fading realizations, the reflection matrix $\mathbf{\Phi}$ is directly correlated with the channel matrices $\mathbf{H}$ and $\mathbf{G}$. 
	\vspace{-0.4cm}
	\subsection{Numerical Analysis}
	As given in Algorithm 1, since the proposed  algorithm relates the phase  shifts of the reflecting elements with the channel statistics, the reflection matrix $\mathbf{\Phi}$ is directly correlated with the channel matrices $\mathbf{H}$ and $\mathbf{G}$. 
	Therefore, the distribution of the composite channel matrix $\mathbf{C}$ could not be derived through a fully analytic approach.  As a result,    we resort to a comprehensive numerical analysis  in order to obtain the statistics of the channel matrix $\mathbf{C}$. 
	
	As it is stated in the previous section, the   elements of the channel  matrices $\mathbf{H}$ and $\mathbf{G}$ are i.i.d. and  follow $\mathcal{CN}(0,1)$ distribution.  Our comprehensive numerical analysis, which performs  $10^6$ Monte-Carlo trials  for each $R_x\times T_x$ configuration, indicates that   the elements of the  composite $\mathbf{C}$ matrix are complex Gaussian random variables with  $\mathcal{CN}(N\mu,N)$ distribution, where $\mu$ is numerically calculated,  in terms of $T_x$ and $R_x$, as
	\begin{equation}
	\mu=\frac{1.8}{(1+2T_x)(1+2R_x)}.
	\label{mu}
	\end{equation}
	
	For supporting the accuracy of this estimation,  in Fig. \ref{fig2},   $(\ref{mu})$ is compared to the the channel statistics of $\mathbf{C}$ for $N=1$, which is obtained through the Monte Carlo simulations  performing  at least $10^6$  trials for each $R_x\times T_x$ set-up. The results  show that  the  estimated $\mu$ in (\ref{mu})  perfectly  fit the  computer simulations per $R_x\times T_x$. 
	
	Then,  (\ref{mu}) is utilized to determine the ABEP of the system in the following subsection. %In order to further support
	\begin{figure}[t]
		\centering
		\includegraphics[width=0.9\linewidth]{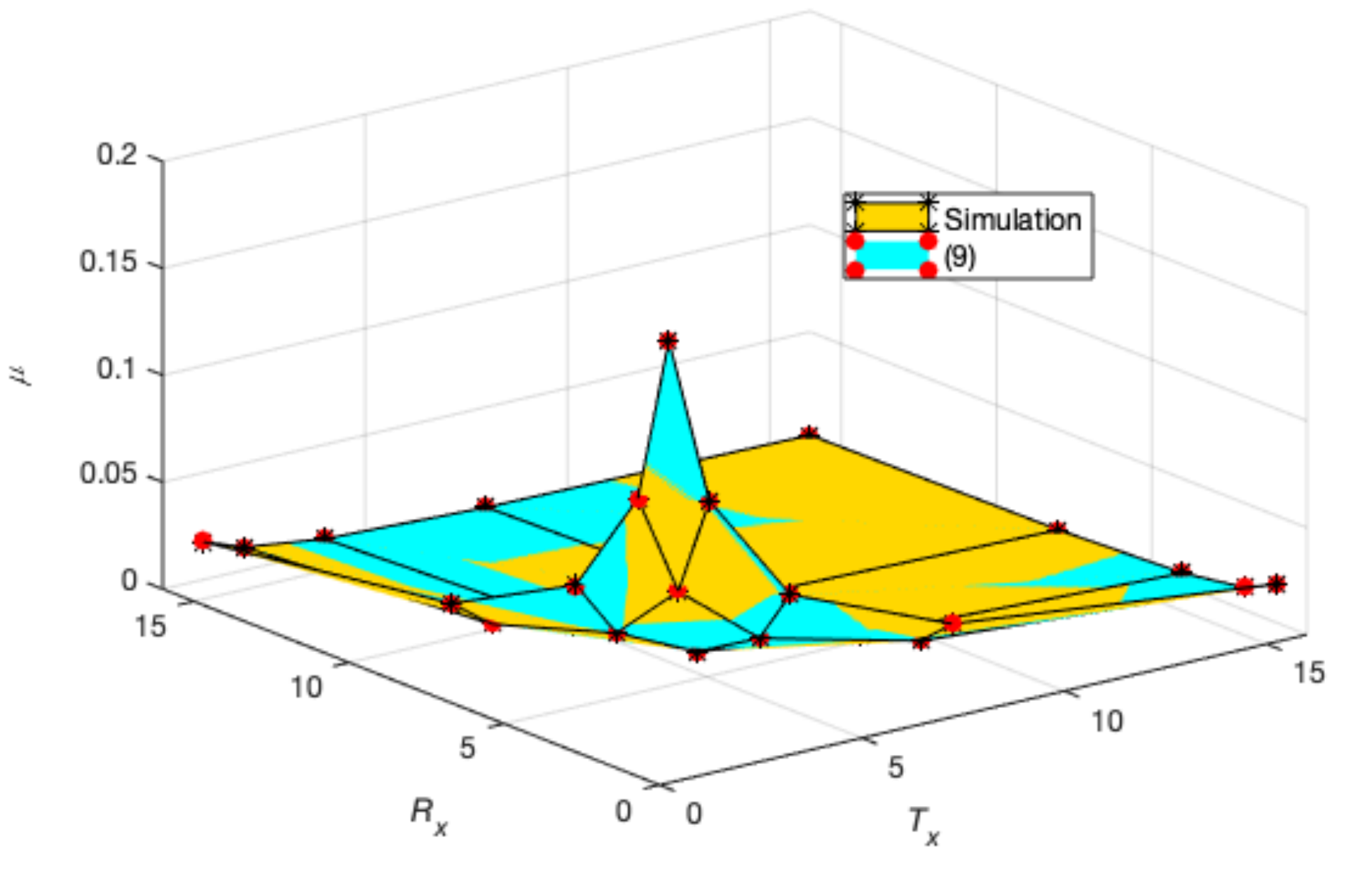}
		\vspace{-0.2cm}
		\caption{Comparison of  the mean $\mu$   (\ref{mu})  with the Monte-Carlo simulations  for different $T_x$ ve $R_x$ antennas.}
		\label{fig2}
			\vspace*{-0.3cm}
	\end{figure}
	\vspace*{-0.35cm}
	\subsection{ABEP Analysis}
	In this subsection, after obtaining a numerical approximation for the statistics of the channel matrix $\mathbf{C}$, an upper bound expression for the ABEP of the proposed system is given as follows \cite{simon2005digital}:
	\begin{equation}
	P_e\leq\frac{1}{\kappa2^\kappa}\sum_{\mathbf{x}}\sum_{\hat{\mathbf{x}}} P(\mathbf{x}\rightarrow\hat{\mathbf{x}})e(\mathbf{x},\hat{\mathbf{x}})
	\label{abep}
	\end{equation}
	where $\kappa$ is the number of incoming information bits, $P(\mathbf{x}\rightarrow \hat{\mathbf{x}})$ is the unconditional pairwise error probability (PEP)  and  $e(\mathbf{x},\hat{\mathbf{x}})$ is the number of error bits for the corresponding PEP event. 
	
	To obtain the PEP expression, first, the conditional PEP (CPEP) of the system is derived,  using the $Q$-function, as follows 
	\begin{equation}
	\vspace*{-0.17cm}
	P(\mathbf{x}\rightarrow \hat{\mathbf{x}}|\mathbf{C})=Q\Bigl(\sqrt  {\frac{\Omega }{2\sigma^2}}\Bigr) 
	\label{cp1}
	\end{equation}
	where $\Omega$ is given,  for $\mathbf{\Delta}=(\mathbf{x}-\hat{\mathbf{x}})(\mathbf{x}-\hat{\mathbf{x}})^H$, {as} %. 
	%Then, for $\mathbf{\Delta}=(\mathbf{x}-\hat{\mathbf{x}})(\mathbf{x}-\hat{\mathbf{x}})^H$, $\Omega$  is  rewritten  in the following quadratic form as
	{\begin{align}
		&\Omega=||\mathbf{C} \big( \mathbf{x}-\hat{\mathbf{x}} \big)||^2\\
		&\hspace{0.25cm}=\mathrm{vec}({\mathbf{C}}^{\mathrm{H}})^{\mathrm{H}}(\mathbf{\Delta}\otimes\mathbf{I}_{R_x})\mathrm{vec}({\mathbf{C}}^{\mathrm{H}}).
		\label{cp2}
		\end{align}
		Therefore,  considering  $Q(x)=\frac{1}{\pi}\int_{0}^{\pi/2}e^{-x^2/2\sin^2\theta}\mathit{d}\theta$, the CPEP   (\ref{cp1}) can be rewritten as
		
		\begin{align}
		\begin{split}
		&P\left( \mathbf{x}\rightarrow\hat{\mathbf{x}}|\mathbf{C}\right)=\\&\hspace{0.5cm}\frac{1}{\pi}\int_{0}^{{\pi/2}}\exp\Bigl( -\varphi\frac{\mathrm{vec}({\mathbf{C}}^{\mathrm{H}})^{\mathrm{H}}(\mathbf{\Delta}\otimes\mathbf{I}_{R_x})\mathrm{vec}({\mathbf{C}}^{\mathrm{H}})}{4\sin^2\theta}  \Bigr) \mathit{d}\theta
		\label{cp3}
		\end{split}
		\end{align}}where $\varphi=1/N_0$.  {Then, averaging (\ref{cp3}) over the matrix $\mathbf{C}$ through moment generating function (MGF) approach results in the following PEP expression
	\begin{align}
	\begin{split}
	&P\left( \mathbf{x}\rightarrow\hat{\mathbf{x}}\right)=\frac{1}{\pi}\int_{0}^{{\pi/2}}M_{\Omega}\biggl( \frac{-\varphi}{4\sin^2\theta}  \biggr) \mathit{d}\theta
	\label{p1}.
	\end{split}
	\end{align}

In \cite{turin1960characteristic}, the MGF of $\mathbf{z}^\mathrm{H}\mathbf{Bz}$, for any Hermitian $\mathbf{B}$ matrix, is given by
	\begin{equation}
	M(s)=\text{exp}\biggl(\frac{s\bar{\mathbf{z}}^\mathrm{H}\mathbf{B}(\mathbf{I}-s\mathbf{C_zB})^{-1}\bar{\mathbf{z}}}{\det(\mathbf{I}-s{\mathbf{C_zB}})}\biggr)
	\label{mgf}
	\end{equation}
	where $\bar{\mathbf{z}}$ is the mean vector and $\mathbf{C_z}$ is the covariance matrix of the vector $\mathbf{z}$.
	
	Clearly, in our system $\mathbf{B}=\mathbf{\Delta}\otimes\mathbf{I}_{R_x}$ and $\mathbf{z}={\vect({\mathbf{C}}^{\mathrm{H}}})$. Using the   numerically obtained channel statistics given in the previous subsection,  $\bar{\mathbf{z}}=\overline{\vect({\mathbf{C}}^{\mathrm{H}})}=\mu N\mathbf{1}$ and   {$\mathbf{C_z}=N\mathbf{I}$} are respectively defined as the mean vector  and the covariance matrix for the Gaussian vector $\mathbf{z}=\vect({\mathbf{C}}^{\mathrm{H}})$.  Therefore, the PEP of the system becomes
	\begin{align}
	\begin{split}
	&P\left( \mathbf{x}\rightarrow\hat{\mathbf{x}}\right)=\\
	&\frac{1}{\pi}\int_{0}^{\pi/2}\frac{\exp\biggr(-\bar{\mathbf{z}}^\mathrm{H}\frac{\varphi}{4\sin^2{\theta}}\mathbf{B}\biggr( \mathbf{I}+\frac{\varphi}{4\sin^2{\theta}}\mathbf{C_z}\mathbf{B}\biggl)^{-1}\bar{\mathbf{z}}\biggl)}{\det\biggl( \mathbf{I}+\frac{\varphi}{4\sin^2\theta}\mathbf{C_z}\mathbf{B}\biggr)}\mathit{d}\theta.
	\label{p2}
	\end{split}
	\end{align}}
\begin{comment}

	Assume the quadratic form expression in (\ref{cp2}) is given in terms of a Hermitian matrix $\mathbf{B}=\mathbf{\Delta}\otimes\mathbf{I}_{R_x}$ and complex Gaussian vector $\mathbf{z}={\vect({\mathbf{C}}^{\mathrm{H}}})$ as $\Omega=\mathbf{z}^{\mathrm{H}}\mathbf{Bz}$. Then, averaging (\ref{cp3}) over the matrix $\mathbf{C}$
	through the moment generating function (MGF) approach of this quadratic form  \cite{turin1960characteristic} results in the following PEP expression
	\begin{align}
	\begin{split}
	&P\left( \mathbf{x}\rightarrow\hat{\mathbf{x}}\right)=\\
	&\frac{1}{\pi}\int_{0}^{\pi/2}\frac{\exp\Bigr(-\bar{\mathbf{z}}^\mathrm{H}\frac{\varphi}{4\sin^2{\theta}}\mathbf{B}\Bigr( \mathbf{I}+\frac{\varphi}{4\sin^2{\theta}}\mathbf{C_z}\mathbf{B}\Bigl)^{-1}\bar{\mathbf{z}}\Bigl)}{\det\bigl( \mathbf{I}+\frac{\varphi}{4\sin^2\theta}\mathbf{C_z}\mathbf{B}\bigr)}\mathit{d}\theta
	\label{p2}
	\end{split}
	\end{align}
	where, given  the numerically obtained channel statistics in the previous subsection, $\bar{\mathbf{z}}=\overline{\vect({\mathbf{C}}^{\mathrm{H}})}=\mu N\mathbf{1}$ and   {$\mathbf{C_z}=N\mathbf{I}$} are respectively defined as the  mean vector  and the covariance matrix for the Gaussian vector $\mathbf{z}=\vect({\mathbf{C}}^{\mathrm{H}})$. 
	content...
\end{comment}
\vspace*{-0.5cm}
	\section{Simulation Results}
	In this section, the BER performance of the proposed RIS-MIMO and RIS-SM schemes are  investigated through theoretical analysis and comprehensive computer simulations. All results are performed as a function of transmitted signal energy to noise ratio ($E_s/N_0$) and for different $R_x\times T_x$ MIMO configurations and BPSK modulation ($M=2$).
		\begin{figure}[t]
		\centering
			\vspace{-0.25cm}
		\includegraphics[width=0.85\linewidth]{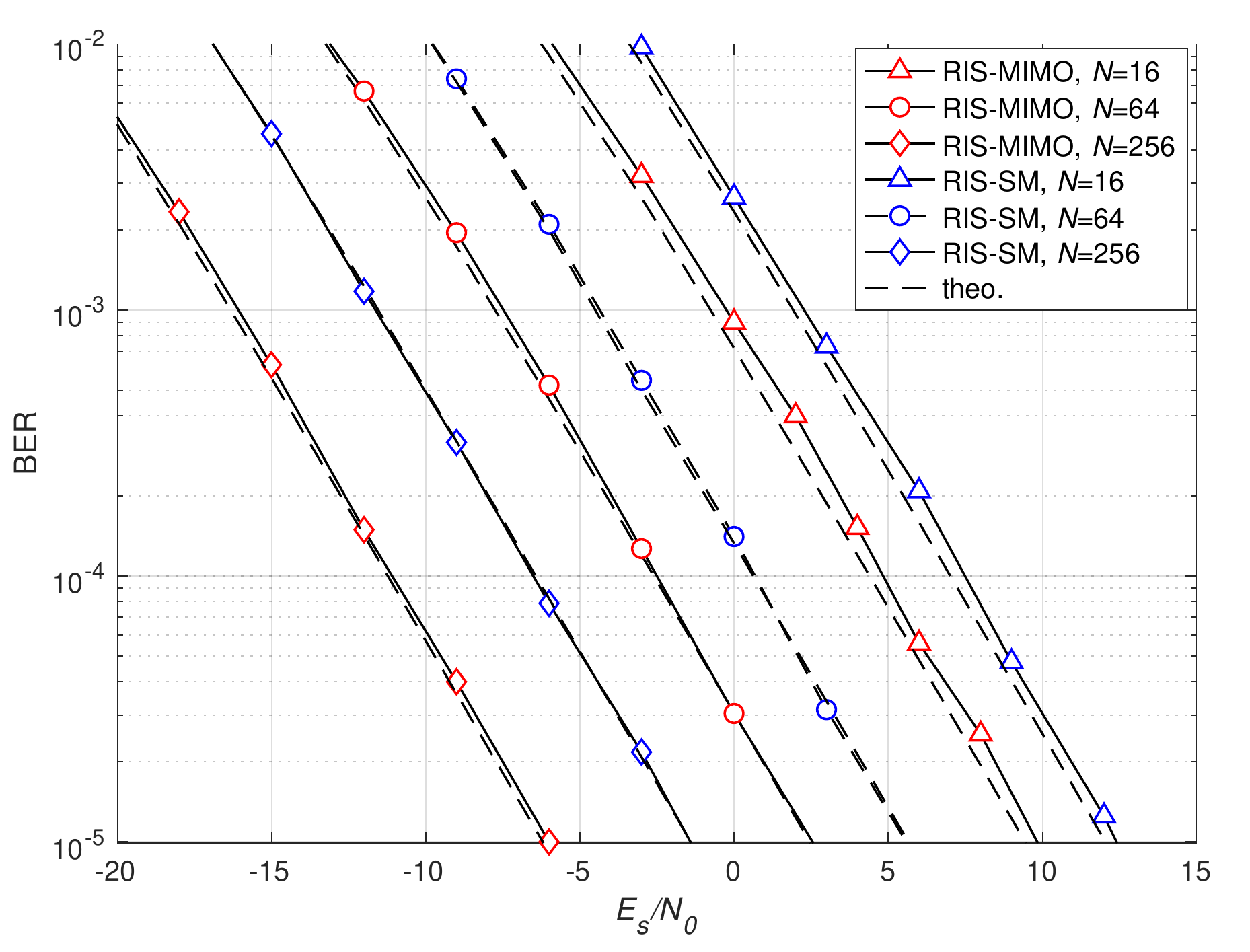}
			\vspace*{-0.4cm}
		\caption{Theoretical and simulation results of the RIS-MIMO ve  RIS-SM schemes.}
		%	\vspace*{-0.1cm}
		\label{fig3}	
	\end{figure}
		\begin{figure}[t]
		\centering
		\vspace{-0.45cm}
		\subfloat[]{{\includegraphics[width=0.45\linewidth]{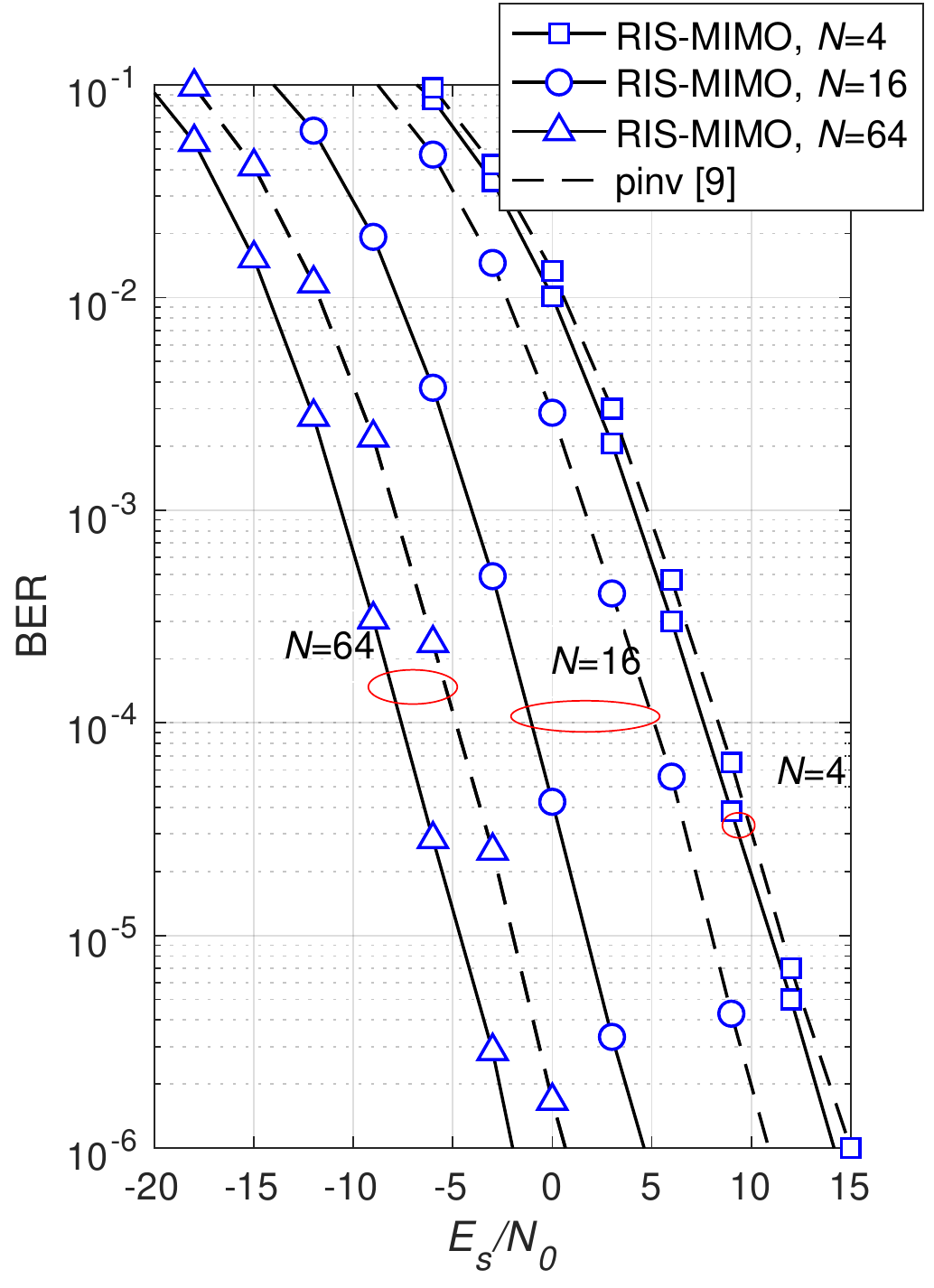} }}%
		%	\hspace{-em}
		\subfloat[]{{\includegraphics[width=0.45\linewidth]{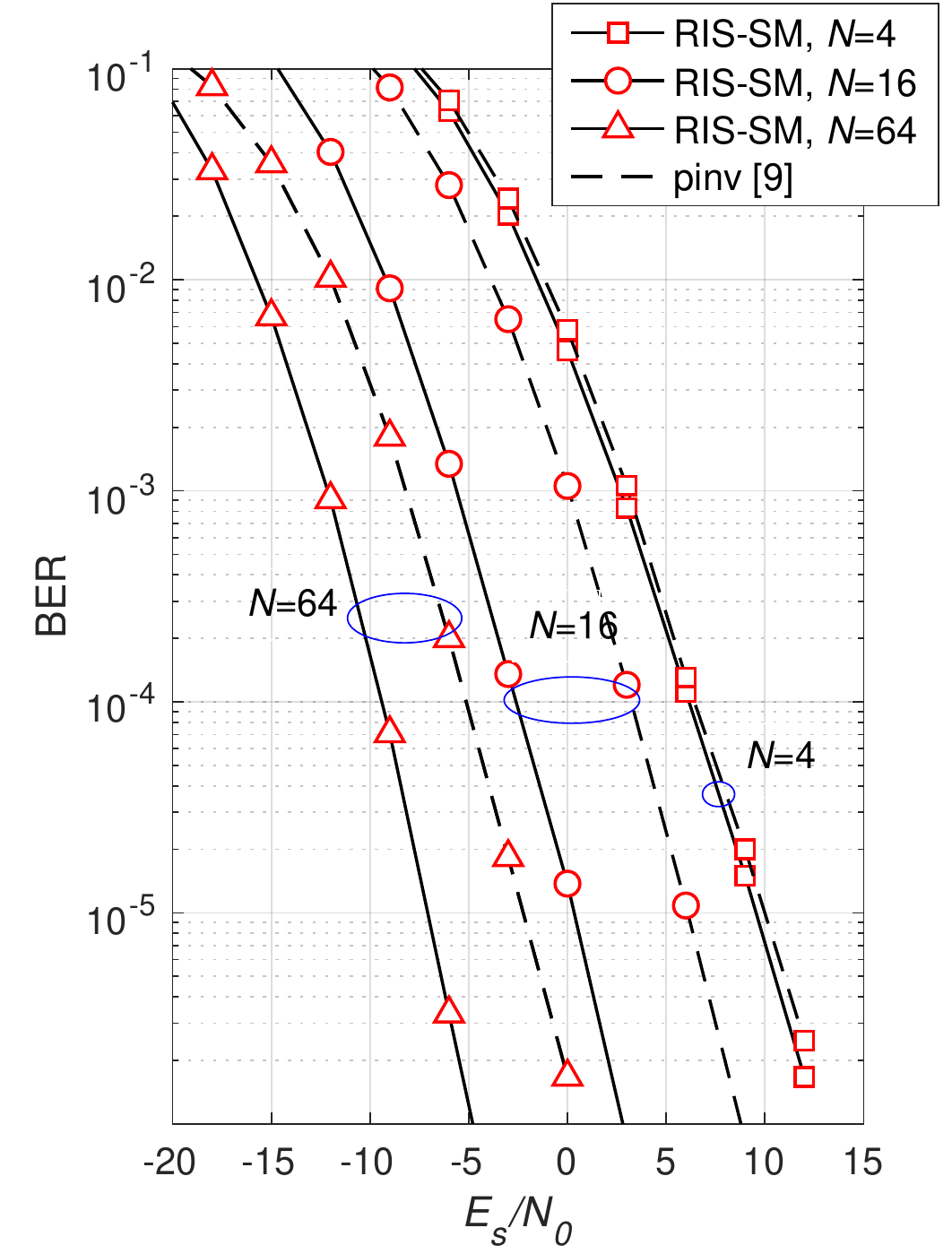} }}%,
		\vspace*{-0.2cm}
		\caption{ {Comparison  of the proposed algorithm   to the pinv algoritm \cite{hou2019mimo}  on the BER performance of the a) RIS-MIMO and b) RIS-SM schemes. }}%
		\label{fig4}%
	\end{figure}

	In Fig. \ref{fig3}, for $4\times 2$ MIMO and various $N$ reflecting elements,   the theoretical BER performance of RIS-MIMO and RIS-SM schemes  are compared with the computer simulation results. It is obvious from this figure that the derived semi-analytic results  perfectly match with the simulation results as $N$ increases.
	
	In Fig. \ref{fig4}, for $4\times 4$ system configuration, the BER performance of the   RIS-MIMO and  RIS-SM schemes using the proposed cosine similarity theorem-based algorithm  and  pseudoinverse (pinv)-based algorithm   \cite{hou2019mimo}  is compared. The results show that compared to   \cite{hou2019mimo},  the BER performance of both RIS-MIMO and RIS-SM {schemes using the proposed algorithm, which requires significantly lower computational complexity, improves  better  as $N$ increases. }
	\begin{figure}[t]
		\centering
		{{\includegraphics[width=0.85\linewidth]{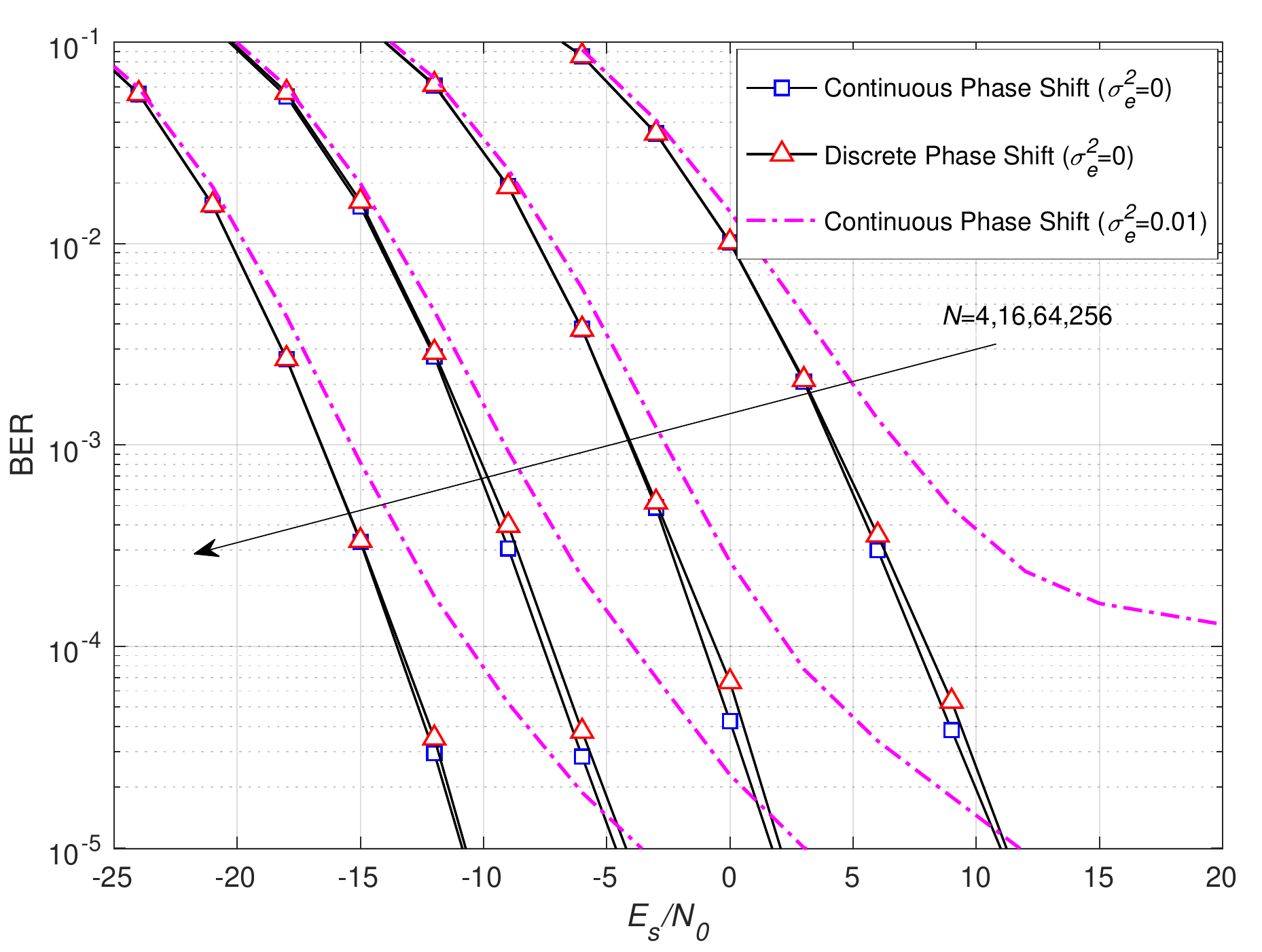} }}%
			\vspace*{-0.35cm}
		\caption{{BER Performance of RIS-MIMO scheme with discrete and continous phase reflection  for the channel estimation error variance  $\sigma_e^2$.  } }%
		\label{fig6}%
		\vspace*{-0.25cm}
	\end{figure}

	{As an illustration, } the proposed algorithm with $\mathcal{O}(6N(T_x+R_x))$   complexity performs $3072$ RMs for   $N=64$,  while the reference pinv algorithm \cite{hou2019mimo} with $\mathcal{O}(N^3)$  complexity performs  $262144$ RMs. This means, the proposed algorithm provides $ 98.8\%$ reduction in computational  complexity  over the reference algorithm \cite{hou2019mimo} for $N=64$. %and at the same time, the proposed algorithm achieves  $3$ ve $5$ dB $E_s/N_0$ gains for RIS-MIMO and RIS-SM schemes, respectively.
	
{	In Fig. 5, the impact of   the imperfect channel knowledge and the use of discrete phase shifts on the BER performance  of the $4\times 4$ RIS-MIMO scheme is investigated for $N$ reflecting elements. } 

{For  discrete phase shifts model, we assume that each reflecting element at the RIS takes a finite number of discrete phase values. Therefore,   a set of discrete phase shifts  is constructed  by uniformly quantizing $[0,\pi]$\footnote{In Algorithm I, for $\arccos(\cdot)$ operation, MATLAB function $\textit{acos}()  $  is used, which returns phase values between the interval $[0,\pi]$.}} { interval into $N$ levels as { $\mathcal{F}=\{0, \:\Delta\phi, \:\ldots,\: (N-1)\Delta\phi\}$, where $\Delta\phi=\pi/N$. Then, the continous phase shift  of the each  reflecting element, which is obtained via the proposed algorithm, is rounded  to the nearest discrete level. The simulation results demonstrate that  discrete phase reflection causes  a minor BER performance degradation compared to the continous phase reflection. However, this performance difference  gradually disapperars as $N$ increases.}}

	{Moreover, to analyze the effect of the channel estimation error,  the imperfect  CSI is assumed at both  the RIS  and  the receiver. Therefore,  we consider that  both the RIS and 	the} {receiver   errorneously estimate $\mathbf{H}$ and $\mathbf{G}$ channel matrices as $\bar{\mathbf{H}}\in\mathbb{C}^{N\times T_x}=\mathbf{H}+\mathbf{E_t}$ and $\bar{\mathbf{G}}\in\mathbb{C}^{N\times R_x}=\mathbf{G}+\mathbf{E_r}$, respectively,  where $\mathbf{E_t}$ and $\mathbf{E_r}$ are the matrices of  the channel estimation errors whose each entry is i.i.d. and  distributed as $\mathcal{CN}(0,\sigma_e^2)$. Therefore, after the proposed algorithm determines the diagonal reflection matrix $\bar{\mathbf{\Phi}}\in\mathbb{C}^{N\times N}$ considering the imperfect channels $\bar{\mathbf{H}}$ and $\bar{\mathbf{G}}$, the overall estimated channel matrix becomes  $\bar{\mathbf{C}}\in\mathbb{C}^{R_x\times T_x}=\bar{\mathbf{G}}^{\mathrm{H}}\bar{\mathbf{\Phi}}\bar{\mathbf{H}}$. %Therefore,  the entries of erroneously estimated channel matrix $\bar{\mathbf{C}}$ are distributed as  $\mathcal{CN}(N\mu, N+\sigma_e^2)$. 
	The simulation results reveal  that  the assumption of  the imperfect CSI  at both the RIS and the receiver  significantly   degrades the BER performance of the RIS-MIMO scheme. However,  increasing the number of reflecting elements at the RIS reduces  this performance degradation.}% the effect of channel errors becomes negligible, i.e., $N+\sigma_e^2\approx N$. Then, the RIS-MIMO system becomes more robust to the  imperfect CSI.}
	
	Fig. \ref{fig5} demonstrates the BER performance of the proposed  RIS-MIMO  scheme  with $4\times 4$ when path loss effect \cite{ellingson2019path} is considered.  Denoting  the distances  from the transmitter to the RIS and from the RIS to the receiver as  $d_1$ and $d_2$, respectively, the path loss $P_L$ of the overall system %from the transmitter to the receiver
	is calculated  as \cite{ellingson2019path}
	\begin{equation}
	P_L^{-1}=\frac{\lambda^4}{256\pi^2}\frac{1}{d^2_1d^2_2} 
	\label{pl}
	\end{equation}
		\begin{figure}[t]
		\vspace{-0.2cm}
		\centering
		{{\includegraphics[width=0.85\linewidth]{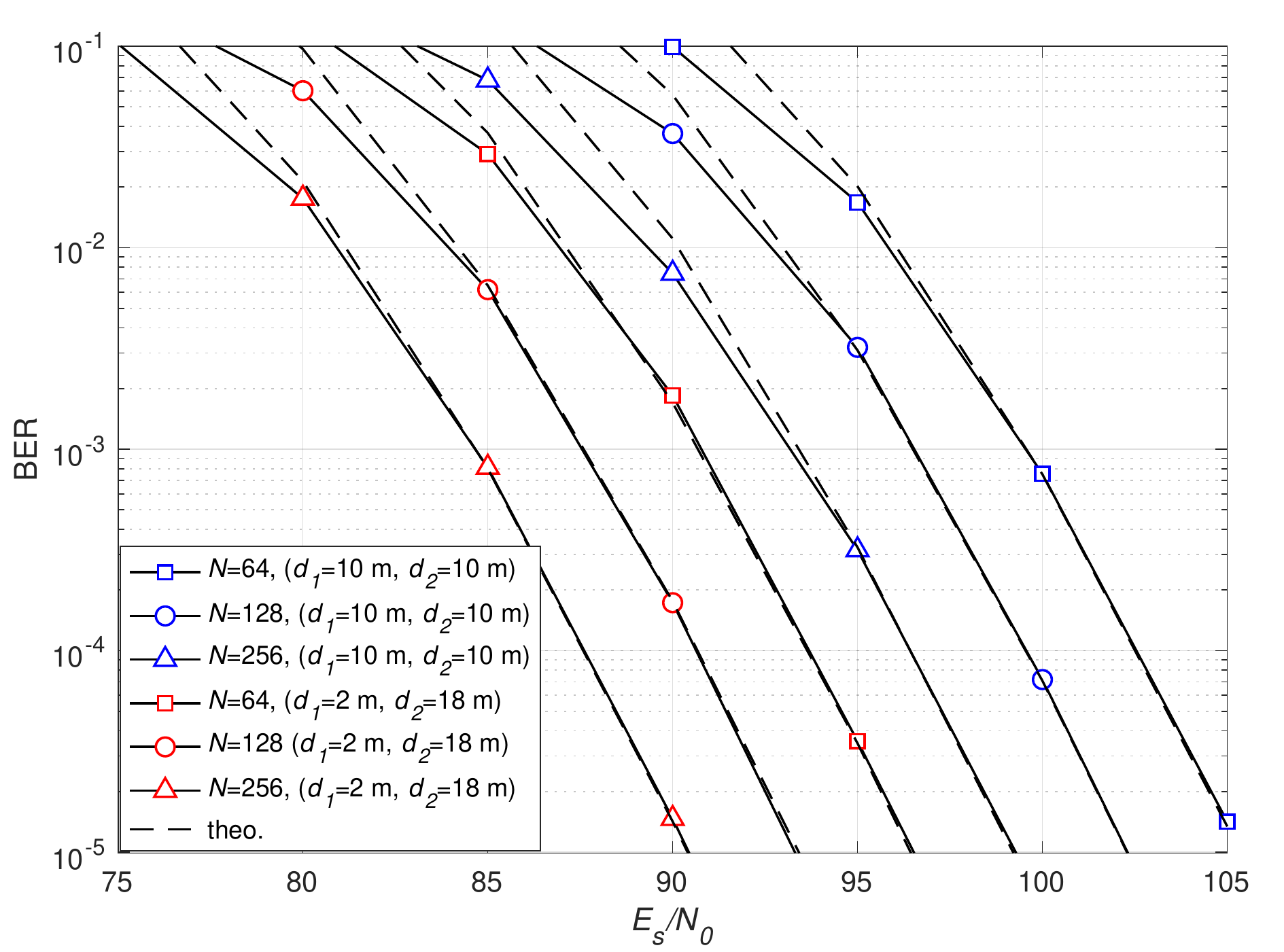} }}%
		\vspace*{-0.35cm}
		\caption{{BER Performance of RIS-MIMO scheme in case of  the path loss \mbox{effect \cite{ellingson2019path}.}} }%
		\label{fig5}%
	\end{figure}where $\lambda$ is the wavelength at $2.4$ GHz operating frequency. The simulations are performed for   $d_1=d_2=10 $ m and $d_1=2$ m, $d_2=18$ m. It can be deduced from the results that  compared to the performance  in the absence of the  path loss given in Fig. \ref{fig4}, 
	the error performance of the RIS-MIMO scheme  significantly degrade with the path loss effect. 
	On the other hand, since the $P_L$ is proportional to $d_1^2d_2^2$,   the BER performance of the system fairly improves when  the RIS is closer to the transmitter or the receiver ($d_1=2$ m   and $d_2=18$ m).
%	\vspace*{-0.3cm}

	\section{Conclusion}
	In this letter,  a   low-complexity algorithm has been developed exploiting the cosine similarity theorem for RIS aided MIMO transmission schemes to enhance the overall path gain of the  communication channel. Moreover, performing a semi-analytic approach, the ABEP of the system has been derived. Furthermore, through the {comprehensive  computer simulations} the BER performance of RIS aided MIMO  schemes {for different circumstances including discrete phase reflection, path loss effect and imperfect CSI    has been investigated}}.  
	\vspace{-0.3cm}
	\bibliographystyle{IEEEtran}
	\bibliography{IEEEabrv,refer}

\begin{thebibliography}{10}
\providecommand{\url}[1]{#1}
\csname url@samestyle\endcsname
\providecommand{\newblock}{\relax}
\providecommand{\bibinfo}[2]{#2}
\providecommand{\BIBentrySTDinterwordspacing}{\spaceskip=0pt\relax}
\providecommand{\BIBentryALTinterwordstretchfactor}{4}
\providecommand{\BIBentryALTinterwordspacing}{\spaceskip=\fontdimen2\font plus
\BIBentryALTinterwordstretchfactor\fontdimen3\font minus
  \fontdimen4\font\relax}
\providecommand{\BIBforeignlanguage}[2]{{%
\expandafter\ifx\csname l@#1\endcsname\relax
\typeout{** WARNING: IEEEtran.bst: No hyphenation pattern has been}%
\typeout{** loaded for the language `#1'. Using the pattern for}%
\typeout{** the default language instead.}%
\else
\language=\csname l@#1\endcsname
\fi
#2}}
\providecommand{\BIBdecl}{\relax}
\BIBdecl

\bibitem{zhang2019capacity}
S.~Zhang and R.~Zhang, ``Capacity characterization for intelligent reflecting
  surface aided {MIMO} communication,'' \emph{arXiv preprint arXiv:1910.01573},
  2019.

\bibitem{nadeem2019large}
Q.-U.-A. Nadeem, A.~Kammoun \emph{et~al.}, ``Large intelligent surface assisted
  {MIMO} communications,'' \emph{arXiv preprint arXiv:1903.08127}, 2019.

\bibitem{basar2019transmission}
E.~Basar, ``Transmission through large intelligent surfaces: A new frontier in
  wireless communications,'' in \emph{IEEE European Conf. Netw. and Commun.
  (EuCNC)}, June 2019, pp. 112--117.

\bibitem{basar2019wireless}
E.~Basar, M.~Di~Renzo \emph{et~al.}, ``Wireless communications through
  reconfigurable intelligent surfaces,'' \emph{IEEE Access}, vol.~7, pp.
  116\,753--116\,773, Aug. 2019.

\bibitem{basar2020reconfigurable}
E.~Basar, ``Reconfigurable intelligent surface-based index modulation: A new
  beyond {MIMO} paradigm for 6{G},'' \emph{IEEE Trans. Commun.}, Feb. 2020.

\bibitem{gopi2020intelligent}
S.~Gopi, S.~Kalyani, and L.~Hanzo, ``Intelligent reflecting surface assisted
  beam index-modulation for millimeter wave communication,'' \emph{arXiv
  preprint arXiv:2003.12049}, 2020.

\bibitem{huang2019reconfigurable}
C.~Huang, A.~Zappone \emph{et~al.}, ``Reconfigurable intelligent surfaces for
  energy efficiency in wireless communication,'' \emph{IEEE Tran. Wireless
  Commun.}, vol.~18, no.~8, pp. 4157--4170, June 2019.

\bibitem{wu2019intelligent}
Q.~Wu and R.~Zhang, ``Intelligent reflecting surface enhanced wireless network
  via joint active and passive beamforming,'' \emph{IEEE Trans. Wireless
  Commun.}, vol.~18, no.~11, pp. 5394--5409, Aug. 2019.

\bibitem{hou2019mimo}
T.~Hou, Y.~Liu \emph{et~al.}, ``{MIMO} assisted networks relying on large
  intelligent surfaces: A stochastic geometry model,'' \emph{arXiv preprint
  arXiv:1910.00959}, 2019.

\bibitem{ellingson2019path}
S.~W. Ellingson, ``Path loss in reconfigurable intelligent surface-enabled
  channels,'' \emph{arXiv preprint arXiv:1912.06759}, 2019.

\bibitem{tang2019wireless}
W.~Tang, M.~Z. Chen \emph{et~al.}, ``Wireless communications with
  reconfigurable intelligent surface: Path loss modeling and experimental
  measurement,'' \emph{arXiv preprint arXiv:1911.05326}, 2019.

\bibitem{basar2020indoor}
E.~Basar, I.~Yildirim, and I.~F. Akyildiz, ``{I}ndoor and outdoor physical
  channel modeling and efficient positioning for reconfigurable intelligent
  surfaces in mm{W}ave bands,'' \emph{arXiv preprint arXiv:2006.02240}, 2020.

\bibitem{basar2020simris}
------, ``Sim{RIS} channel simulator for reconfigurable intelligent
  surface-empowered communication systems,'' \emph{arXiv preprint
  arXiv:2006.00468}, 2020.

\bibitem{abeywickrama2020intelligent}
S.~Abeywickrama, R.~Zhang \emph{et~al.}, ``Intelligent reflecting surface:
  Practical phase shift model and beamforming optimization,'' \emph{arXiv
  preprint arXiv:2002.10112}, 2020.

\bibitem{wu2019weighted}
Q.~Wu and R.~Zhang, ``Weighted sum power maximization for intelligent
  reflecting surface aided {SWIPT},'' \emph{IEEE Wireless Commun. Lett.},
  vol.~9, no.~5, pp. 586--590, 2019.

\bibitem{moon2000mathematical}
\color{black}T. K.~Moon and W.~C. Stirling, \emph{\color{black}Mathematical
  {M}ethods and {A}lgorithms for {S}ignal {P}rocessing}.\hskip 1em plus 0.5em
  minus 0.4em\relax Prentice Hall Upper Saddle River, NJ, 2000, vol.~1.

\bibitem{mesleh2008spatial}
R.~Y. Mesleh, H.~Haas \emph{et~al.}, ``Spatial modulation,'' \emph{IEEE Trans.
  Veh. Technol.}, vol.~57, no.~4, pp. 2228--2241, July 2008.

\bibitem{simon2005digital}
M.~K. Simon and M.-S. Alouini, \emph{Digital communication over fading
  channels}.\hskip 1em plus 0.5em minus 0.4em\relax John Wiley \& Sons, 2005,
  vol.~95.

\bibitem{turin1960characteristic}
G.~L. Turin, ``The characteristic function of hermitian quadratic forms in
  complex normal variables,'' \emph{Biometrika}, vol.~47, no. 1/2, pp.
  199--201, 1960.

\end{thebibliography}

\end{document}